\documentclass[reqno]{amsart}
\usepackage{amsmath, amssymb, amscd}

\theoremstyle{plain}

\newtheorem{prop}{Proposition}[section]
\newtheorem{thm}{Theorem}[section]

\theoremstyle{definition}

\theoremstyle{remark}

\newcommand{\alphabr}{{\overset{\curvearrowright}{\boldsymbol{\alpha}}}_k}
\newcommand{\alphabl}{{\overset{\curvearrowleft}{\boldsymbol{\alpha}}}_k}
\newcommand{\alphal}[1][k]{{\overset{\curvearrowleft}{\alpha}}_{#1}}
\newcommand{\alphar}[1][k]{{\overset{\curvearrowright}{\alpha}}_{#1}}
\newcommand{\ovr}[1][x]{\overset{\curvearrowright}{#1}}
\newcommand{\ovl}[1][x]{\overset{\curvearrowleft}{#1}}

\begin{document}
\numberwithin{equation}{section}

\title{Superspinors}
\author{Alexander Golubev}
\address{New York Institute of Technology, New York, New York}
\email{agolubev@nyit.edu}
\begin{abstract}
We propose to replace $\text{Spin}(1,3)^e$ as the  space-time symmetry 
group of quantum field theory by a compact semisimple Lie group. 
The results are rendered via the formalism of superspinors - objects
identifiable as particle or antiparticle wave functions,
and governed by the Fermi-Dirac statistics. 
\end{abstract}

\maketitle

\section{Introduction}
In this paper we attempt to replace the Lorentz group of space-time symmetries
with a compact semisimple Lie group of symmetries of purely quantum objects. 
The rationale behind such an attempt is deceptively simple:
the quantum field theoretical data ought to transform unitarily. Admittedly,
our approach is not the only one. There are unitary representations
of the classical Lorentz group, and if the technical difficulties posed by
their infinite dimensionality are overcome, there would appear to be no call
for replacing the group. Be that as it may, there is another compelling 
reason to look for a different symmetry group. The seeming disparity in the way 
electrons and positrons are treated~\cite{D} has to be either explained in 
terms of fundamental space-time symmetries or done away with, 
and the present physical setup does not do that.\\
\indent
There were several attempts in the past. The most notable and fruitful
replacement candidate had been the conformal group. Even though those 
transformations only leave the light cone intact, while wrecking havoc 
on the time-like dynamics, such achievements as conformal field theory, 
the Penrose transform and the formalism of twistors in the curved space-time 
~\cite{P},~\cite{P-R} had validated that particular break from the grip of 
the Lorentz group, as well as inspired further research.\\
\indent
Whatever the motivation, this replacement ushers in some new features, 
akin to supersymmetric theories, and has experimentally verifiable 
consequences. Instead of considering Dirac spinors ${\Psi}$ and ${\Phi}$
delineating particles and antiparticles as separate entities, we unify them.
Mathematically, this unification is expressed by the formalism of superspinors 
- objects appearing to different observers as particle or antiparticle wave
functions, depending on a particular frame of reference (parameterized 
by the frame rapidity $\alpha$): 
\begin{equation*}
{\Psi}(\alpha + \pi)= \pm \bar{\Phi}(\alpha).
\end{equation*} 
\noindent
To elicit these transformations, we no longer require the standard Dirac
Lagrangian:
\begin{equation*}
{\mathcal{L}}_D = \frac{i}{2}({\Psi}^{\dag}{\gamma}^{\mu}
{\partial}_{\mu}{\Psi}- {\partial}_{\mu}{\Psi}^{\dag}{\gamma}^{\mu}
{\Psi}- m{\Psi}^{\dag}{\Psi}).
\end{equation*}
\noindent
Instead, we introduce a modified Lagrangian: 
\begin{equation*}
{\mathcal{L}}_D = \frac{i}{2}({\Psi}^{\dag}{\gamma}^{\mu}
{\nabla}_{\mu}(\alpha){\Psi}- 
{\nabla}_{\mu}(\alpha){\Psi}^{\dag}{\gamma}^{\mu}
{\Psi}- m{\Psi}^{\dag}{\Psi}),
\end{equation*}
\noindent
where ${\nabla}_{\mu}(\alpha)$ is a family of principal connections
subjected to the relativistic constraint
\begin{equation*}
g^{\nu \eta}(\alpha){\nabla}_{\nu}(\alpha){\nabla}_{\eta}(\alpha)=
{\partial}_{\mu}{\partial}^{\mu}, 
\end{equation*}
\noindent
and the metric $g^{\nu \eta}(\alpha)$ is induced by the 
deformations germane to the new group.\\ 
\indent
Another curious feature of the new group is that superspinors
are fermions \textit{par excellence}. That is to say, in the course
of second quantization some appropriate anticommutators vanish
as a result of fairly natural assumptions.\\
\indent
Novelties notwithstanding, the rotation properties of spinors go
over to superspinors, owing to the $SU(2)$ subgroup common
to both the Lorentz group and the new one. Therefore, superspinors
in a rest frame coincide with the Dirac spinors.\\ 
\indent
A few words about the paper. Its organization is straightforward:
first we develop the necessary Lie group theory in Sections 2 and 3, 
then to make it usable we modify the concept of free spin structure 
in Section 4, and finally Sections 5, 6, 7 contain some application 
of the aforementioned mathematics to the spinorial representations.
Section 8 is not as rigorous, an homage to the experimental aspects 
of superspinors.\\ 
\indent
Lastly, we dispense with the physical constants by setting
$ \hbar \;=\;c\;=\;1.$

%\newpage

\section{Compactification of the symmetry group}

\indent
The point of departure for a symmetry group search is the consideration 
of the oriented Grassmanian manifold ${\mathfrak{Gr^{+}}}^3_6$
of 3-planes in ${\mathbb{R}}^6$ as a natural arena to tackle
the inertial frames of ${\mathbb{R}}^4$.
Just a glimpse of what we are up against. 
A boost in the $x$ direction is given by
\begin{equation}
x' = \frac{x +vt}{\sqrt{1 - v^2}}, \quad t' = \frac{t +vx}{\sqrt{1 - v^2}}.
\end{equation}
\noindent
As $v \to 1$, we have
\begin{equation}\label{singularity}
\lim_{v \to 1}\left(\arcsin \left(\frac{vt}{\sqrt{x^2 + v^2t^2}}\right)- 
\arcsin \left(\frac{t}{\sqrt{t^2 +v^2x^2}}\right)\right) = 0.
\end{equation}
\noindent
Not only are there infinite lengths (which can be easily normalized away), 
but also the frames as such cease to exist at $v=1$.
As demonstrated by~\eqref{singularity}, the $x$ and $t$ axes merge.
That necessitates a representation of the Lorentz frames by points of some
projective variety. Our choice (to be justified in due course) is
${\mathfrak{Gr^{+}}}^3_6$.\\
\indent
Fortuitously, ${\mathfrak{Gr^{+}}}^3_6
\subset {\mathbb{S}}^{19}$ (the latter being the unit sphere in 
${\mathbb{R}}^{20}$). This allows us to use the Pl\H{u}cker coordinates 
(and the quadratic Pl\H{u}cker relations since ${\mathfrak{Gr^{+}}}^3_6$
is a proper subvariety of ${\mathbb{S}}^{19}$). For a comprehensive reference on
the Pl\H{u}cker coordinates, see the classic by Hodge and Pedoe (\cite{Hodge},
Chapter VII). 
Thus given a $3 \times 6$ matrix of rank 3, there are precisely 20 $3 \times 3$
minors (not counting the column permutations). Their determinants are not all
zero because of the rank condition, and comprise the set of Pl\H{u}cker
coordinates of the 3-plane spanned by the row vectors. These are unique up to
a common positive multiple. We denote them by $p_{i_0 i_1 i_2}$, where 
${i_{\lambda}}$'s are distinct numbers from the set $(0, 1, 2, 3, 4, 5)$ with 
the additional property $i_0< i_1< i_2$. We arrange the 20 
$p_{i_0 i_1 i_2}$'s in lexicographic order. The aggregate of these entities
can be thought of as a surjective mapping $\mathsf{P}$ from the set of all 
$3\times 6$ matrices of rank 3 onto the Grassmanian.\\
\indent
A point $P \in {\mathfrak{Gr^{+}}}^3_6$ is completely determined by its coordinates: 
\begin{equation*}
P = (p_{012}, p_{013}, p_{014}, .........., p_{345}).
\end{equation*} 
\noindent
Obversely, a list of $p_{i_0 i_1 i_2}$'s
does not designate a point unless they are antisymmetric in all their
indices and satisfy the quadratic relations 
\begin{equation}
F_{i_0 i_1 j_0 j_1 j_2 j_3 }(P)=0,
\end{equation}
\noindent
where by definition, 
\begin{equation}\label{plucker}
F_{i_0 i_1 j_0 j_1 j_2 j_3 }(P)= \sum_{\lambda =0}^3
(-1)^{\lambda}p_{i_0 i_1 j_{\lambda}}p_{j_0... j_{\lambda -1} j_{\lambda +1}
...j_3}.
\end{equation}
\noindent
Both $i$'s and $j$'s are some distinct numbers from the set $(0, 1, 2, 3, 4, 5)$.\\
\indent
We fix a point $P_0 \in {\mathfrak{Gr^{+}}}^3_6$, and choose the Pl\H{u}cker
coordinates so that
$$P_0 = (1, 0, 0, 0, 0, 0, 0, 0, 0, 0, 0, 0, 0, 0, 0, 0,  0, 0, 0, 0).$$
These coordinates can be gleaned off the matrix
\begin{equation}\label{initial matrix}
\begin{bmatrix}
1 & 0& 0& 0& 0& 0\\
0&  1& 0& 0& 0& 0\\
0&  0& 1& 0& 0& 0
\end{bmatrix}.
\end{equation}
\noindent
Corresponding to this matrix is the standard orthonormal frame of 
${\mathbb{R}}^6$ which we denote by $(x^1, x^2, x^3, t^3, t^2, t^1)$. 
On this 6-dimensional space, we first fix a $t$, $t$ being an 
orthogonal linear combination of $t^l$'s. Without loss of generality 
we may set $t =t^3$. Then to each value of the boost parameters
\begin{equation}
{\alphar} = \arctan \tanh \left(\frac{\frac{\partial x^k}{\partial t}}
{\sqrt{1 -\left({\frac{\partial x^k}{\partial t}}\right)^2}} \right),
\quad {\alphar} \in [0, \frac{\pi}{4}), \quad k= \{1,2,3\},
\end{equation}
\noindent
we associate a point $P({\alphar}) \in {\mathfrak{Gr}}^3_6$ via
\begin{equation}\label{alpha1}
\begin{bmatrix}
\cos 4 \alphar[1] & 0& 0&\sin 4 \alphar[1] & 0 & 0 \\
0 &1 & 0 &0 & 0 & 0  \\
0& 0& 1& 0& 0& 0\\
-\sin 4 \alphar[1] & 0& 0& \cos 4 \alphar[1] &0 &0\\
0& 0& 0&0 &1&0\\
0& 0&0& 0& 0& 1
\end{bmatrix},
\end{equation}
\begin{equation}\label{alpha2}
\begin{bmatrix}
1& 0 & 0 & 0 & 0 & 0 \\
0 &\cos 4 \alphar[2] & 0 &\sin 4 \alphar[2] &0&0  \\
0& 0&1& 0&0 &0\\
0& -\sin 4 \alphar[2] &0&\cos 4 \alphar[2]  &0 &0\\
0& 0&0&0 &1&0\\
0&0&0& 0& 0&1 
\end{bmatrix},
\end{equation}
\begin{equation}\label{alpha3}
\begin{bmatrix}
1& 0 & 0 & 0 & 0 & 0 \\
0 &1& 0 &0&0&0  \\
0& 0& \cos 4 \alphar[3] &\sin 4 \alphar[3]  &0 &0\\
0& 0&-\sin 4 \alphar[3] &\cos 4 \alphar[3]  &0 &0\\
0& 0&0&0 &1&0\\
0&0&0& 0& 0&1 
\end{bmatrix}
\end{equation}
\noindent
followed by $\mathsf{P}$. 
Thereafter we denote~\eqref{alpha1}-\eqref{alpha3} by 
$\alphabr$, and the image points of individual 
${\mathsf{P}} \circ {\alphabr}$ by $P({\alphar})$. \\
\indent
If instead of~\eqref{initial matrix} we
choose some other matrix of rank 3, yielding the same Pl\H{u}cker coordinates, 
the latter would be related to~\eqref{initial matrix} by a matrix 
$A \in GL(3, \mathbb{R})^e$ so that the span of row space would remain unchanged.
Therefore~\eqref{alpha1}-\eqref{alpha3} are independent of the choice 
of~\eqref{initial matrix}. Furthermore, the diagram below would commute
(`${\times |A|}$'  means scalar multiplication by $\det A$).
\begin{equation*}\begin{CD}
{\alphabr}@>{\mathsf{P}}>>{\mathsf{P}} \circ {\alphabr}\\
@V{A}VV              @VV{\times |A|}V\\
{A{\alphabr}} @>{\mathsf{P}}>>{\mathsf{P}} \circ (A{\alphabr}) 
\end{CD}\end{equation*} 
\noindent
Such a swap will always preserve the resulting Pl\H{u}cker coordinates. \\ 
\indent
At this point we state and prove an important theorem regarding the
properties of~\eqref{alpha1}-\eqref{alpha3}.
\begin{thm}\label{thm1} 
The mappings 
\begin{equation*}
\mathsf{P} \circ {\alphabr}: \;\;\; 
{\mathbb{R}}^+ \hookrightarrow {\mathfrak{Gr^{+}}}^3_6
\end{equation*}
effected by the composition of $\mathsf{P}$ and
~\eqref{alpha1}-\eqref{alpha3} are analytic embeddings.
\end{thm}
\begin{proof}
We prove the theorem for a particular case $(k=1)$.
The remaining cases would then follow \textit{mutatis mutandis}. 
From~\eqref{alpha1} we construct the transformation matrix:
\begin{equation}
\begin{bmatrix}
\cos 4 \alphar[1] & 0& 0&\sin 4 \alphar[1] & 0 & 0 \\
0& 1& 0& 0& 0& 0\\
0& 0& 1& 0& 0& 0
\end{bmatrix},
\end{equation}
The list of all nonzero (for some $\alphar$) Pl\H{u}cker coordinates is as follows:
\begin{equation}
\begin{array}{lllll}
p_{012}= {\cos} 4 \alphar[1],\\ 
p_{123}= \sin 4 \alphar[1].
\end{array}
\end{equation}
From this, the analyticity is immediate. Demonstrating the one-to-oneness
is slightly more involved. For the time being we think of 
$\mathsf{P} \circ {\alphabr}$ as a mapping into ${\mathbb{S}}^{19}$. 
This is a 19-dimensional analytic manifold. On this manifold, there is a 
natural system of charts $(V_{p_{i_0 i_1 i_2}}, z_{m_1 m_2 m_3})$ such that
$V_{p_{i_0 i_1 i_2}}$ is the family of hypersurfaces with 
${p_{i_0 i_1 i_2}} \ne 0$, analytically diffeomorphic to an open 
subset of ${\mathbb{R}}^{19}$ coordinatized by
\begin{equation*}
z_{m_1 m_2 m_3} = \frac{p_{m_1 m_2 m_3}}{p_{i_0 i_1 i_2}},\;\;\;
{m_1 m_2 m_3} \ne {\sigma(i_0) \sigma(i_1) \sigma(i_2)}\;\;\;
\forall \sigma.
\end{equation*}
\noindent
In our case, ${p_{i_0 i_1 i_2}} = p_{012}$ or 
${p_{i_0 i_1 i_2}} = p_{123}$. The image is contained in $V_{p_{012}}$ 
and $V_{p_{123}}$. Hence the existence of a smooth nonvanishing tangent 
vector to our image curve would suffice. The respective non-zero components 
of the tangent vector are:
\begin{equation*}
z'_{123}=\frac{-4}{{\sin}^2 4 \alphar[1]} 
\quad \text{within}\; V_{p_{012}},
\end{equation*}
\begin{equation*}
z'_{012}=\frac{4}{{\cos}^2 4 \alphar[1]} 
\quad \text{within}\; V_{p_{123}}.
\end{equation*}
Their being smooth and nonvanishing clinches the proof.
\end{proof}
\noindent
We use the notation ${LGr}_{\curvearrowright}$ to name the aggregate
image of $\mathsf{P} \circ {\alphabr}, \; k= \{1,2,3\}$. The
geometric meaning of ${LGr}_{\curvearrowright}$ is transparent. Each
point represents a 3-plane transversal to the natural foliation 
of ${\mathbb{R}}^4$ by the hyperbolic hypersurfaces parameterized via 
($t^2 -(x^{k_1})^2 = a> 0, \ x^{k_2},\ x^{k_3}$), and intersecting every 
leaf of that foliation. All the planes generated by a boost in any 
particular direction assembled would reconstitute the inside of the
corresponding light wedge $t^2 - (x^k)^2 =0$. We are as yet to establish 
a link between our construct and the classical Lorentz group. Our claim 
is, there is such a link, and, in fact, the set of all spatially
rotated boost frames $SO(3)\cdot {LGr}_{\curvearrowright}$
accounts for all orthochrone boosts. Let us take a look at the 
$SO(1,3)^e/SO(3)$ bundle over $SO(3)\cdot{LGr}_{\curvearrowright}$.  
Define a canonical section of this bundle by
\begin{equation}\label{canonical section}
\mathcal{S}: \quad  P(O\cdot\alphar) \longmapsto B(O \cdot\alphar).
\end{equation}
\noindent
Here $B(\alphar)$ are the standard $4 \times 4$
matrices representing the boosts of $SO(1,3)^e$. 
This section is very nice. We have
\begin{prop}\label{prop1}
$\mathcal{S}$ defined by~\eqref{canonical section},
\begin{equation*}
\mathcal{S}: \quad SO(3)\cdot {LGr}_{\curvearrowright}
\longrightarrow  SO(1,3)^e/SO(3), \quad \text{is a diffeomorphism.}
\end{equation*}
\end{prop}
\begin{proof}
This is just an elementary application of the Cartan's 
`technique of the graph'~(\cite{C}, for modern treatment 
see~\cite{G}, Lecture 6). We show that the diagonal subset 
\begin{equation*}
\Delta = \{ \alphar \mid \; 
P(O\cdot \alphar)) \times  B(O \cdot\alphar) \}
\end{equation*}
projects diffeomorphically onto the base and the fiber. 
The system of charts used
in the proof of Theorem~\ref{thm1}, from the standard
Euclidean metric, induces an analytic Riemannian metric on 
${\mathbb{S}}^{19}$, \textit{ergo} on ${\mathfrak{Gr^{+}}}^3_6$. 
Employing this metric and the tangent vectors obtained in the 
course of proving Theorem~\ref{thm1}, we get global dual forms 
$d \alphar$. Those are analytic and invariant with respect to the
action of ${\mathsf{P}} \circ {\alphabr}$. Next we take a
right-invariant coframe ${\Omega}_j$ on $SO(1, 3)/SO(3)$. The
exterior differential system
\begin{equation}
{\Delta}^*=
{\pi}^*_{\text{base}}d \ovr[\alpha] -
{\pi}^*_{\text{fiber}} \Omega
\end{equation}
is completely integrable and defines an analytic
foliation of the diagonal subset. To see the injectivity
of ${\pi}_{\text{base}*}$, we assume ${\pi}_{\text{base}*}(X)=0$,
for some vector field tangent to the foliation. Then
\begin{equation}
0 = i_X {\Delta}^* = i_X {\pi}^*_{\text{base}}(\omega, 
d \ovr[\alpha]) - i_X {\pi}^*_{\text{fiber}} \Omega
= - i_X {\pi}^*_{\text{fiber}} \Omega.
\end{equation}
But $\Omega$ is a full coframe, hence ${\pi}_{\text{fiber}*}(X)=0$,
and $X=0$. Since the dimensions of the foliation and the base are
the same and ${\pi}_{\text{base}*}$ is injective, it is an
isomorphism. Now we apply the Inverse Function Theorem to deduce
that the restriction of ${\pi}_{\text{base}*}$ to every leaf is
a local diffeomorphism that happens to be invariant under the
group action on the right. Therefore it is a global diffeomorphism.  
In a similar vein we deal with ${\pi}_{\text{fiber}*}$. By uniqueness, 
$\Delta$ is the graph of  $\mathcal{S}$, and the proposition now follows. 
\end{proof}
${LGr}_{\curvearrowright}$  is not closed in the quotient topology of 
${\mathfrak{Gr^{+}}}^3_6$. Now we manufacture the set 
$\overline{{LGr}_{\curvearrowright}}$. To be able 
to adjoin the limiting points, we have to check if they are 
\textit{bona fide} elements of the Grassmanian. This amounts to 
verifying these two relations: the easy one-
\begin{equation}
\lim_{{\alphar} \to \frac{\pi}{4}} 
{p_{\sigma (i_0) \sigma (i_1) \sigma (i_2)}}
= \text{sign}\; \sigma \lim_{{\alphar} \to \frac{\pi}{4}} 
{p_{i_0 i_1 i_2}},
\end{equation}
and the cumbersome one -~\eqref{plucker};
\begin{equation}
\lim_{{\alphar} \to \frac{\pi}{4}} 
F_{i_0 i_1  j_0 j_1 j_2 j_3 }(P({\alphar}))=0.
\end{equation}
\noindent
Luckily for us, due to the paucity
of nonzero ${p_{i_0 i_1 i_2}}$'s, there is only one nontrivial
identity (disregarding index permutations) for each 
transformation~\eqref{alpha1}-\eqref{alpha3}. We have
\begin{gather}
\begin{split}
\lim_{{\alphar[1]} \to \frac{\pi}{4}} 
F_{120123}(P({\alphar[1]})) &= \lim_{{\alphar[1]} \to \frac{\pi}{4}} 
(p_{012}p_{123} -p_{123}p_{012})\\
&= \lim_{{\alphar[1]} \to \frac{\pi}{4}} 
(\cos 4\alphar[1] \sin 4\alphar[1]-
\sin 4\alphar[1] \cos 4\alphar[1])\\
& = 0 \quad \text{ for~\eqref{alpha1}},
\end{split}\\
\begin{split}
\lim_{{\alphar[2]} \to \frac{\pi}{4}} 
F_{020123}(P({\alphar[2]}))&= \lim_{{\alphar[2]} \to \frac{\pi}{4}} 
(p_{012}p_{023} -p_{023}p_{012})\\
&= \lim_{{\alphar[2]} \to \frac{\pi}{4}} 
(\cos 4\alphar[2](- \sin 4\alphar[2])-
(-\sin 4\alphar[2]) \cos 4\alphar[2])\\
&= 0 \quad \text{ for~\eqref{alpha2}},
\end{split}\\
\begin{split}
\lim_{{\alphar[3]} \to \frac{\pi}{4}} 
F_{010123}(P({\alphar[3]}))&= \lim_{{\alphar[3]} \to \frac{\pi}{4}} 
(p_{012}p_{013} -p_{013}p_{012})\\
&= \lim_{{\alphar[3]} \to \frac{\pi}{4}} 
(\cos 4\alphar[3] \sin 4\alphar[3]-
\sin 4\alphar[3] \cos 4\alphar[3])\\
&= 0 \quad \text{ for~\eqref{alpha3}}.
\end{split}
\end{gather}
\noindent
For all three transformations, there is just one limiting point:
\begin{equation*}
\lim_{{\alphar} \to \frac{\pi}{4}}  P(\alphar)= P_{\infty}=
(-1, 0, 0, 0, 0, 0, 0, 0, 0, 0, 0, 0, 0, 0, 0, 0,  0, 0, 0, 0).
\end{equation*}
\indent
We have succeeded in building $\overline{{LGr}_{\curvearrowright}}
= {LGr}_{\curvearrowright} \cup P_{\infty}$. But a larger question
is still looming: how to complete $\overline{{LGr}_{\curvearrowright}}$
to a group space for some one-parameter subgroups of $SO(6)$?
Evidently we need more Lorentz boost frames. The problem is, according
to Proposition~\ref{prop1}, all those frames are represented by the 
points of ${LGr}_{\curvearrowright}$. The frames in the other 
component of $SO(1, 3)$ are essentially parachrone boosts and cannot 
be connected to $SO(1, 3)^e$ (or any representation of it) via
continuous transformations. The same is true of the remaining
two components of the classical Lorentz group. To assuage this
deficiency, we introduce the notion of `virtual frame'. Within our
realm we represent frames by the points of ${\mathfrak{Gr^{+}}}^3_6$,
i. e. by the appropriately positioned 3-planes in ${\mathbb{R}}^6$.
Now to complete $\overline{{LGr}_{\curvearrowright}}$ we use the 
symmetry properties of ${\mathfrak{Gr^{+}}}^3_6$. In keeping with the
physical world, we supply a more concrete
description. Thus, the `virtual frames' correspond to the situation
wherein 3-planes assembled would reconstitute the outside of the 
light wedge $t^2 - (x^k)^2 = 0$. Going from a `real frame' to a 
`virtual frame'
amounts to flipping the signature of the Lorentzian metric involved. 
From our viewpoint, the virtuality is manifested in the parameters
being reciprocal to those of~\eqref{alpha1}-\eqref{alpha3}:\\
\begin{equation}
{\alphal} = \arctan \tanh \left(\frac{\frac{\partial t}{\partial x^k}}
{\sqrt{1 -\left({\frac{\partial t}{\partial x^k}}\right)^2}} \right),
\quad {\alphal} \in [0, \frac{\pi}{4}), \quad k= \{1,2,3\}.
\end{equation}
\noindent
The corresponding transformations are listed below:\\ 
\begin{equation*}\tag{\ref{alpha1}$'$}\label{alpha4}
\begin{bmatrix}
-\cos 4\alphal[1] & 0& 0&-\sin 4\alphal[1] & 0 & 0 \\
0 &1 & 0 &0 & 0 & 0  \\
0& 0& 1& 0& 0& 0\\
\sin 4\alphal[1] & 0& 0& -\cos 4\alphal[1] &0 &0\\
0& 0& 0&0 &1&0\\
0& 0&0& 0& 0& 1
\end{bmatrix},
\end{equation*}
\begin{equation}\tag{\ref{alpha2}$'$}\label{alpha5}
\begin{bmatrix}
1& 0 & 0 & 0 & 0 & 0 \\
0 &-\cos 4 \alphal[2] & 0 &-\sin 4 \alphal[2] &0&0  \\
0& 0&1& 0&0 &0\\
0& \sin 4 \alphal[2] &0&-\cos 4 \alphal[2]  &0 &0\\
0& 0&0&0 &1&0\\
0&0&0& 0& 0&1 
\end{bmatrix},
\end{equation}
\begin{equation*}\tag{\ref{alpha3}$'$}\label{alpha6}
\begin{bmatrix}
1& 0 & 0 & 0 & 0 & 0 \\
0 &1& 0 &0&0&0  \\
0& 0& -\cos 4 \alphal[3] &-\sin 4 \alphal[3]  &0 &0\\
0& 0&\sin 4 \alphal[3] &-\cos 4 \alphal[3]  &0 &0\\
0& 0&0&0 &1&0\\
0&0&0& 0& 0&1 
\end{bmatrix}.
\end{equation*}
\noindent
Just as before, the above transformations possess the expected
properties.
\begin{thm}\label{thm2} 
The mappings 
\begin{equation*}
\mathsf{P} \circ {\alphabl}: \;\;\; 
{\mathbb{R}}^+ \hookrightarrow {\mathfrak{Gr^{+}}}^3_6
\end{equation*}
effected by the composition of $\mathsf{P}$ and
~\eqref{alpha4}-\eqref{alpha6} are analytic embeddings.
\end{thm}
\noindent
The limiting process works as well:
\begin{equation*}
\lim_{{\alphal} \to \frac{\pi}{4}}  P(\alphal)= P_0=
(1, 0, 0, 0, 0, 0, 0, 0, 0, 0, 0, 0, 0, 0, 0, 0,  0, 0, 0, 0).
\end{equation*}
\noindent
The sets ${LGr}_{\curvearrowright}$ and ${LGr}_{\curvearrowleft}$ 
are not connected so that the common boundary set is
\begin{equation}\label{boundset}
\overline{{LGr}_{\curvearrowright}} \cap
\overline{{LGr}_{\curvearrowleft}} = \{ P_0, P_{\infty}\}.
\end{equation}
\noindent
Now we act on $\overline{{LGr}_{\curvearrowright}}$ by spatial
rotations. The resulting augmented set is
\begin{equation}
(SO(3) \cdot \overline{{LGr}_{\curvearrowright}}) \subset
{\mathfrak{Gr^{+}}}^3_6. 
\end{equation}
\noindent
Similarly, we obtain
$(SO(3) \cdot \overline{{LGr}_{\curvearrowleft}}) \subset
{\mathfrak{Gr^{+}}}^3_6$. An essential relation holding
true for those sets is that
\begin{equation}\label{intersection}
(SO(3) \cdot \overline{{LGr}_{\curvearrowright}}) \cap
(SO(3) \cdot \overline{{LGr}_{\curvearrowleft}})= 
\{ P_0, P_{\infty}\}.
\end{equation}
\noindent
As it turns out, ${\mathfrak{Gr^{+}}}^l_6$ is the 
lowest-dimensional projective space with enough room to
accommodate~\eqref{intersection}. That is possible only if
\begin{equation}
\dim {\mathfrak{Gr^{+}}}^l_m = \binom{m}{l}-1 \geqslant 12.
\end{equation}
\noindent
From this one readily sees that $m \geqslant 6$.
By contrast, the common boundary of $(SO(3) \cdot 
\overline{{LGr}_{\curvearrowright}})$ and 
$(SO(3) \cdot \overline{{LGr}_{\curvearrowleft}})$ embedded in
${\mathfrak{Gr^{+}}}^3_4$ would have been homeomorphic to 
${\mathbb{S}}^2$, and because of it being connected there would
be ways to move from $P_0$ to $P_{\infty}$ via spatial rotations.\\
\indent
An important consequence of Theorem~\ref{thm1}, Theorem~\ref{thm2},
and~\eqref{boundset} is the following statement:
\begin{thm}
\begin{equation*}
H^1 (\overline{{LGr}_{\curvearrowright}} \cup
\overline{{LGr}_{\curvearrowleft}}, \mathbb{Z}) \cong
{\mathbb{Z}}^{15}.
\end{equation*}
\end{thm} 
\indent
Now we are in a position to unveil the Lie algebra
$\overset{\curvearrowright}{\mathfrak{g}} \cong \mathfrak{so}(1,3)$
underpinning \eqref{alpha1}-\eqref{alpha3} and \eqref{alpha4}-\eqref{alpha6}
(which from this point on are parametrized by ${\alpha}_k$). 
To begin with, we express the boosts in
terms of the standard orthogonal Lie algebra basis:
\begin{equation}
{\ovr[K]}_1 =
\begin{bmatrix}
\phantom{-}0&\phantom{-}0&\phantom{-}0&\phantom{-}i&\phantom{-}0&\phantom{-}0\\
\phantom{-}0&\phantom{-}0&\phantom{-}0&\phantom{-}0&\phantom{-}0&\phantom{-}0\\
\phantom{-}0&\phantom{-}0&\phantom{-}0&\phantom{-}0&\phantom{-}0&\phantom{-}0\\
-i          &\phantom{-}0&\phantom{-}0&\phantom{-}0&\phantom{-}0&\phantom{-}0\\
\phantom{-}0&\phantom{-}0&\phantom{-}0&\phantom{-}0&\phantom{-}0&\phantom{-}0\\
\phantom{-}0&\phantom{-}0&\phantom{-}0&\phantom{-}0&\phantom{-}0&\phantom{-}0
\end{bmatrix},
\end{equation}
\begin{equation}
{\ovr[K]}_2 =
\begin{bmatrix}
\phantom{-}0&\phantom{-}0&\phantom{-}0&\phantom{-}0&\phantom{-}0&\phantom{-}0\\
\phantom{-}0&\phantom{-}0&\phantom{-}0&\phantom{-}i&\phantom{-}0&\phantom{-}0\\
\phantom{-}0&\phantom{-}0&\phantom{-}0&\phantom{-}0&\phantom{-}0&\phantom{-}0\\
\phantom{-}0&-i          &\phantom{-}0&\phantom{-}0&\phantom{-}0&\phantom{-}0\\
\phantom{-}0&\phantom{-}0&\phantom{-}0&\phantom{-}0&\phantom{-}0&\phantom{-}0\\
\phantom{-}0&\phantom{-}0&\phantom{-}0&\phantom{-}0&\phantom{-}0&\phantom{-}0
\end{bmatrix},
\end{equation}
\begin{equation}
{\ovr[K]}_3 =
\begin{bmatrix}
\phantom{-}0&\phantom{-}0&\phantom{-}0&\phantom{-}0&\phantom{-}0&\phantom{-}0\\
\phantom{-}0&\phantom{-}0&\phantom{-}0&\phantom{-}0&\phantom{-}0&\phantom{-}0\\
\phantom{-}0&\phantom{-}0&\phantom{-}0&\phantom{-}i&\phantom{-}0&\phantom{-}0\\
\phantom{-}0&\phantom{-}0&-i          &\phantom{-}0&\phantom{-}0&\phantom{-}0\\
\phantom{-}0&\phantom{-}0&\phantom{-}0&\phantom{-}0&\phantom{-}0&\phantom{-}0\\
\phantom{-}0&\phantom{-}0&\phantom{-}0&\phantom{-}0&\phantom{-}0&\phantom{-}0
\end{bmatrix}.
\end{equation}
\noindent
Their brackets 
\begin{equation}
[{\ovr[K]}_1,{\ovr[K]}_2 ] =i{\ovr[J]}_3,
\quad[{\ovr[K]}_2,{\ovr[K]}_3 ] =i{\ovr[J]}_1,
\quad[{\ovr[K]}_3,{\ovr[K]}_1 ] =i{\ovr[J]}_2,
\end{equation}
\noindent
yield the rotations:
\begin{equation}
{\ovr[J]}_1 =
\begin{bmatrix}
\phantom{-}0&\phantom{-}0&\phantom{-}0&\phantom{-}0&\phantom{-}0&\phantom{-}0\\
\phantom{-}0&\phantom{-}0&\phantom{-}i&\phantom{-}0&\phantom{-}0&\phantom{-}0\\
\phantom{-}0&          -i&\phantom{-}0&\phantom{-}0&\phantom{-}0&\phantom{-}0\\
\phantom{-}0&\phantom{-}0&\phantom{-}0&\phantom{-}0&\phantom{-}0&\phantom{-}0\\
\phantom{-}0&\phantom{-}0&\phantom{-}0&\phantom{-}0&\phantom{-}0&\phantom{-}0\\
\phantom{-}0&\phantom{-}0&\phantom{-}0&\phantom{-}0&\phantom{-}0&\phantom{-}0

\end{bmatrix},
\end{equation}
\begin{equation}
{\ovr[J]}_2 =
\begin{bmatrix}
\phantom{-}0&\phantom{-}0&\phantom{-}i&\phantom{-}0&\phantom{-}0&\phantom{-}0\\
\phantom{-}0&\phantom{-}0&\phantom{-}0&\phantom{-}0&\phantom{-}0&\phantom{-}0\\
-i          &\phantom{-}0&\phantom{-}0&\phantom{-}0&\phantom{-}0&\phantom{-}0\\
\phantom{-}0&\phantom{-}0&\phantom{-}0&\phantom{-}0&\phantom{-}0&\phantom{-}0\\
\phantom{-}0&\phantom{-}0&\phantom{-}0&\phantom{-}0&\phantom{-}0&\phantom{-}0\\
\phantom{-}0&\phantom{-}0&\phantom{-}0&\phantom{-}0&\phantom{-}0&\phantom{-}0
\end{bmatrix},
\end{equation}
\begin{equation}
{\ovr[J]}_3 =
\begin{bmatrix}
\phantom{-}0&\phantom{-}i&\phantom{-}0&\phantom{-}0&\phantom{-}0&\phantom{-}0\\
-i          &\phantom{-}0&\phantom{-}0&\phantom{-}0&\phantom{-}0&\phantom{-}0\\
\phantom{-}0&\phantom{-}0&\phantom{-}0&\phantom{-}0&\phantom{-}0&\phantom{-}0\\
\phantom{-}0&\phantom{-}0&\phantom{-}0&\phantom{-}0&\phantom{-}0&\phantom{-}0\\
\phantom{-}0&\phantom{-}0&\phantom{-}0&\phantom{-}0&\phantom{-}0&\phantom{-}0\\
\phantom{-}0&\phantom{-}0&\phantom{-}0&\phantom{-}0&\phantom{-}0&\phantom{-}0
\end{bmatrix}.
\end{equation}
\noindent
The algebra generated by $\ovr[K]$'s and $\ovr[J]$'s is closed:
\begin{equation}
\begin{aligned}
&[{\ovr[J]}_1,{\ovr[J]}_2] = i{\ovr[J]}_3,
\quad [{\ovr[J]}_2,{\ovr[J]}_3 ] =i{\ovr[J]}_1,
\quad [{\ovr[J]}_3,{\ovr[J]}_1 ] =i{\ovr[J]}_2,\\
&[{\ovr[J]}_1,{\ovr[K]}_2 ] = i{\ovr[K]}_3,
\quad [{\ovr[J]}_1,{\ovr[K]}_3 ] =-i{\ovr[K]}_2,
\quad [{\ovr[J]}_2,{\ovr[K]}_1 ] =i{\ovr[K]}_3,\\
&[{\ovr[J]}_2,{\ovr[K]}_3 ] = i{\ovr[K]}_1,
\quad [{\ovr[J]}_3,{\ovr[K]}_1 ] =i{\ovr[K]}_2,
\quad [{\ovr[J]}_3,{\ovr[K]}_2 ] =-i{\ovr[K]}_1,
\end{aligned}
\end{equation} 
and all the remaining brackets vanish.\\
\indent
The upshot of our discourse is that compactification must involve the
adjoining of virtual frames. Indeed, the parameters of the classical
Lorentz group run through the set of nonnegative real numbers; this 
set is not bounded, therefore no point identification or creation of a
compact group space is possible prior to taking some kind of closure.
But once the virtual frames are in, we are forced to treat them just
as we would the inertial frames. In particular, an observer situated
inside would have no means to decide whether their frame is real
or virtual. Consequently, all the foregoing constructing may start off
with the virtual frames as a foundation. That way one obtains an
alternative algebra denoted $\ovl[\mathfrak{g}]$ instead of 
$\ovr[\mathfrak{g}]$. The two are isomorphic but nonetheless not 
identical. We have
\begin{equation}
{\ovl[K]}_3 = {\ovr[K]}_3,
\end{equation}
\begin{equation}
{\ovl[K]}_2 =
\begin{bmatrix}
\phantom{-}0&\phantom{-}0&\phantom{-}0&\phantom{-}0&\phantom{-}0&\phantom{-}0\\
\phantom{-}0&\phantom{-}0&\phantom{-}0&\phantom{-}0&\phantom{-}0&\phantom{-}0\\
\phantom{-}0&\phantom{-}0&\phantom{-}0&\phantom{-}0&\phantom{-}i&\phantom{-}0\\
\phantom{-}0&\phantom{-}0&\phantom{-}0&\phantom{-}0&\phantom{-}0&\phantom{-}0\\
\phantom{-}0&\phantom{-}0& -i         &\phantom{-}0&\phantom{-}0&\phantom{-}0\\
\phantom{-}0&\phantom{-}0&\phantom{-}0&\phantom{-}0&\phantom{-}0&\phantom{-}0
\end{bmatrix},
\end{equation}
\begin{equation}
{\ovl[K]}_1 =
\begin{bmatrix}
\phantom{-}0&\phantom{-}0&\phantom{-}0&\phantom{-}0&\phantom{-}0&\phantom{-}0\\
\phantom{-}0&\phantom{-}0&\phantom{-}0&\phantom{-}0&\phantom{-}0&\phantom{-}0\\
\phantom{-}0&\phantom{-}0&\phantom{-}0&\phantom{-}0&\phantom{-}0&\phantom{-}i\\
\phantom{-}0&\phantom{-}0&\phantom{-}0&\phantom{-}0&\phantom{-}0&\phantom{-}0\\
\phantom{-}0&\phantom{-}0&\phantom{-}0&\phantom{-}0&\phantom{-}0&\phantom{-}0\\
\phantom{-}0&\phantom{-}0& -i         &\phantom{-}0&\phantom{-}0&\phantom{-}0
\end{bmatrix}.
\end{equation}
\noindent
Proceeding as before, we build $\ovl[\mathfrak{g}] \subset \mathfrak{so}(6)$, 
such that $\ovl[\mathfrak{g}] \cong \mathfrak{so}(1, 3)$. 
$\ovr[\mathfrak{g}]$ and $\ovl[\mathfrak{g}]$ are isomorphic, and 
$[{\ovr[J]}_i, {\ovl[J]}_j] =0$ for all pairs $(i, j)$. Those 
commutative brackets enable us to define a new entity - one
that is completely invariant with respect to the vantage point
change -  
$\ovr[\mathfrak{g}] \; {\Join}  \;\ovl[\mathfrak{g}] \subset 
\mathfrak{so}(6)$. Our joint algebra is a closed subalgebra of 
$\mathfrak{so}(6)$, 
$\dim \ovr[\mathfrak{g}] \; {\Join}  \;\ovl[\mathfrak{g}] = 6$. 
Its elements are generated by $ J_i = {\ovr[J]}_i + {\ovl[J]}_i$ 
and $K_3 = {\ovr[K]}_3 = {\ovl[K]}_3$. In view of 
$\ovr[\mathfrak{g}] \cap \ovl[\mathfrak{g}] = \mathbb{R} K_3$,
$\ovr[\mathfrak{g}] \; {\Join}  \;\ovl[\mathfrak{g}] \ne
\ovr[\mathfrak{g}] \; \oplus  \;\ovl[\mathfrak{g}]$ - a nuance
figuring prominently in the following sections. According 
to Helgason (\cite{He}, Chapter~II, \S 2, Theorem~2.1), there is a unique  
connected Lie subgroup of $SO(6)$, whose Lie algebra is the subalgebra 
$\ovr[\mathfrak{g}] \; {\Join}  \;\ovl[\mathfrak{g}]$ 
of $ \mathfrak{so}(6)$. Furthermore, by a fundamental
result of Mostow~\cite{M}, any semisimple Lie subgroup $H$ of a compact 
Lie group $C$ is closed in the relative topology of $C$. In our case,
$SO(6)$ is compact, 
$\ovr[\mathfrak{g}] \; {\Join}  \;\ovl[\mathfrak{g}]$ is semisimple, 
so that the Mostow's theorem applies. Thus we finally obtain
\begin{equation}\label{groupdef}
G \overset{\text{def}}{=}\; \{ \exp i X\;|X \in \ovr[\mathfrak{g}] 
\; {\Join}\;\ovl[\mathfrak{g}] \}.
\end{equation}
\noindent
The group herein defined by~\eqref{groupdef} ought to
replace the classical Lorentz group as the symmetry group of quantum
objects - the only objects surmised to be capable of being virtual. 
\section{Unitary conversion}
\indent
Having thus determined the structure of the group we begin to look 
for an appropriate 
spinor representation of $G$. We utilize a well-known isomorphism of 
Lie algebras. Specifically, $\mathfrak{so}(6) \cong \mathfrak{su}(4)$.      
Via the above isomorphism, we find a closed subalgebra 
$\ovr[\mathfrak{ug}] \; {\Join}  \;\ovl[\mathfrak{ug}]
\subset \mathfrak{su}(4)$,
\begin{equation}\label{uggisomorphism} 
\ovr[\mathfrak{ug}] \; {\Join}  \;\ovl[\mathfrak{ug}] \cong 
\ovr[\mathfrak{g}] \; {\Join}  \;\ovl[\mathfrak{g}].
\end{equation} 
\noindent
Once more invoking (\cite{He}, Chapter II, \S 2, Theorem~2.1), 
and~\cite{M}, we get a closed subgroup of $SU(4)$. This subgroup, 
denoted $UG$, is the unitary counterpart of $G$. The basis of
$\ovr[\mathfrak{ug}] \; {\Join}  \;\ovl[\mathfrak{ug}]$ (with the
notation retained from Section 2) is
\begin{equation}
\begin{array}{cccc}
J_1 = \frac{1}{2}
\begin{bmatrix}
\phantom{-}0&\phantom{-}1&\phantom{-}0&\phantom{-}0\\
 -1         &\phantom{-}0&\phantom{-}0&\phantom{-}0\\
\phantom{-}0&\phantom{-}0&\phantom{-}0&\phantom{-}1\\
\phantom{-}0&\phantom{-}0& -1         &\phantom{-}0
\end{bmatrix},\quad &
J_2 = \frac{1}{2}
\begin{bmatrix}
\phantom{-}0&\phantom{-}i&\phantom{-}0&\phantom{-}0\\
\phantom{-}i&\phantom{-}0&\phantom{-}0&\phantom{-}0\\
\phantom{-}0&\phantom{-}0&\phantom{-}0&\phantom{-}i\\
\phantom{-}0&\phantom{-}0&\phantom{-}i&\phantom{-}0
\end{bmatrix},\\ 
J_3 = \frac{1}{2}
\begin{bmatrix}
\phantom{-}i&\phantom{-}0&\phantom{-}0&\phantom{-}0\\
\phantom{-}0& -i         &\phantom{-}0&\phantom{-}0\\
\phantom{-}0&\phantom{-}0&\phantom{-}i&\phantom{-}0\\
\phantom{-}0&\phantom{-}0&\phantom{-}0&-i
\end{bmatrix},\quad &
K_3 = \frac{1}{2}
\begin{bmatrix}
\phantom{-}0&\phantom{-}0&          -1&\phantom{-}0\\
\phantom{-}0&\phantom{-}0&\phantom{-}0&\phantom{-}1\\
\phantom{-}1&\phantom{-}0&\phantom{-}0&\phantom{-}0\\
\phantom{-}0&          -1&\phantom{-}0&\phantom{-}0
\end{bmatrix},\\   
K_2 = \frac{1}{2}
\begin{bmatrix}
\phantom{-}0&\phantom{-}0&\phantom{-}0&\phantom{-}1\\
\phantom{-}0&\phantom{-}0&\phantom{-}1&\phantom{-}0\\
\phantom{-}0& -1         &\phantom{-}0&\phantom{-}0\\
      -1    &\phantom{-}0&\phantom{-}0&\phantom{-}0
\end{bmatrix},\quad &
K_1 = \frac{1}{2}
\begin{bmatrix}
\phantom{-}0&\phantom{-}0&\phantom{-}0&\phantom{-}i\\
\phantom{-}0&\phantom{-}0&          -i&\phantom{-}0\\
\phantom{-}0&          -i&\phantom{-}0&\phantom{-}0\\
\phantom{-}i&\phantom{-}0&\phantom{-}0&\phantom{-}0
\end{bmatrix}.
\end{array}
\end{equation}
\noindent
$\ovr[\mathfrak{ug}] \; {\Join}  \;\ovl[\mathfrak{ug}]$ 
decomposes as a vector space into two three-dimensional
subspaces,
\begin{equation}\label{orthogonaldecomposition}
\ovr[\mathfrak{ug}] \; {\Join}  \;\ovl[\mathfrak{ug}]
= \mathfrak{j} \oplus \mathfrak{k},
\end{equation}
\noindent 
of which $\mathfrak{j}$ is a closed compactly embedded Lie
subalgebra generated by $J_i$, and the following identities 
hold:
\begin{equation}
[\mathfrak{j}, \mathfrak{k}] \subset \mathfrak{k}, \quad
[\mathfrak{k}, \mathfrak{k}] \subset \mathfrak{j}.
\end{equation}
\noindent
The corresponding group 
\begin{equation}
SU_J(2) \overset{\text{def}}{=} \{\exp iJ \;|J \in 
\mathfrak{j} \subset 
\ovr[\mathfrak{ug}] \; {\Join}  \;\ovl[\mathfrak{ug}]\}
\end{equation}
\noindent
is a closed subgroup of $UG$.\\
\indent
The most significant property of $UG$ is that it serves as a  
covering space for $G$. Formally we have
\begin{thm}
There exists a map $\Xi$, such that
\begin{equation*}
\Xi : \quad UG \longrightarrow G
\end{equation*}
is a twofold covering epimorphism of Lie groups.
\end{thm}
\begin{proof}
We know that the group $SU_J(2)$ is a twofold cover for
$SO(3)\subset G$. That covering property may be
expressed by
\begin{equation*}\left.
\begin{matrix}
\phantom{-} I_{UG}\\
- I_{UG}
\end{matrix}\right\} 
\overset{{\Xi}_{SU_J(2)}}{\longmapsto} I_G.
\end{equation*}
\noindent
From~\eqref{uggisomorphism} and (\cite{He}, Chapter II, \S 1,
Theorem 1.11) we extract a local isomorphism between $UG$ and $G$.
That means there are sets $O_{UG}^{+}$ and $O_{UG}^{-}$, open in
$UG$, $I_{UG} \in O_{UG}^{+}$, $-I_{UG} \in O_{UG}^{-}$, satisfying
$ O_{UG}^{+} \cap O_{UG}^{-} = \emptyset$, and local diffeomorphisms
$f^{+}$ and $f^{-}$, such that 
$f^{+}(O_{UG}^{+}) = f^{-}(O_{UG}^{-}) = O_G$, the latter set being
an open neighborhood of identity in $G$. Shrinking $O_{UG}^{+}$ if
necessary, we find an open neighborhood of identity in $UG$ which 
we call $O$, $O \subset O_{UG}^{+}$ that is particularly amenable 
to the group multiplication on the left. Namely, 
$(-I_{UG} \cdot O)\subset O_{UG}^{-}$. Because of the group structure, 
we have
\begin{equation*}
(u\cdot O)\cap(u\cdot(-I_{UG}\cdot O))=\emptyset,\quad \forall u \in UG.
\end{equation*}
\indent
Now $UG$ is compact and has no small subgroups, that is, given an open
set $T \subset UG$, such that the diameter of $T$ with respect to the
natural left-invariant Killing metric, $\text{diam}(T) < \text{diam}(O)$,
there exists an element $u_T \in O$, and a nonnegative integer $n_T$
with the property $T \subset (u_T^{n_T} \cdot O)$.\\
\indent
We define $\Xi$ by extending $f^{+}$:
\begin{equation*}
{\Xi}(u_T^{n_T} \cdot O)\overset{\text{def}}{=}
(f^{+}(u_T))^{n_T}f^{+}(O).
\end{equation*}
\noindent
The group operation on the right is differentiable, in fact analytic,
so $\Xi$ is differentiable. We claim that $\Xi$ is a twofold covering
map. If $T \subset O^{-}_{UG}$, there are $u_T,\;n_T$ suct that
$u_T^{n_T} = - I_{UG}$, $(f^{+}(u_T))^{n_T} = I_{G}$, so that
$\Xi (T) \subset f^{+}(O)$, and $G$ is evenly covered for any particular
set of $u_T$'s and $n_T$'s.
\end{proof}
\indent
In the sequel we will work with the homogeneous space $UG/SU_J(2)$. Its
topology turns out to be crucial in our efforts to put $UG$ on solid
ground.
\begin{thm}\label{trivialhomotopy}
\begin{equation*}
{\pi}_1(UG/SU_J(2))= 0.
\end{equation*}
\end{thm}
\begin{proof}
For all Lie groups ${\pi}_2 (.) = 0$~\cite{B}; for $SU_J(2)$, 
${\pi}_0(SU_J(2))= 0$ by connectedness. Also, $SU_J(2)$ is a
closed subgroup of $SU(4)$ in the ordinary matrix topology.
We therefore have the following exact homotopy sequence~\cite{B}:
\begin{equation*}
0 \rightarrow {\pi}_2 (SU(4)/SU_J(2)) \rightarrow {\pi}_1 (SU_J(2))
\rightarrow {\pi}_1 (SU(4)) \rightarrow {\pi}_1 (SU(4)/SU_J(2))
\rightarrow 0.
\end{equation*}
\noindent
${\pi}_1 (SU(4))= 0$~\cite{B} whence 
\begin{equation*}
{\pi}_1 (SU(4)/SU_J(2)) \cong {\pi}_1 (SU_J(2)) = 
{\pi}_1({\mathbb{S}}^3) =0.
\end{equation*}
\noindent
Now homotopy is functorial. The embedding
$\xi : UG/SU_J(2) \hookrightarrow SU(4)/SU_J(2)$ induces the
monomorphism of fundamental groups
\begin{equation*}
{\xi}_{\pi *} : {\pi}_1 (UG/SU_J(2)) \rightarrow
{\pi}_1 (SU(4)/SU_J(2)). \qedhere
\end{equation*}
\end{proof}
\begin{thm}
\begin{equation*}\label{s3theorem}
UG/SU_J(2) \cong {\mathbb{S}}^3.
\end{equation*}
\end{thm}
\begin{proof}
Based on the decomposition~\eqref{orthogonaldecomposition},
there is an involutive automorphism 
\begin{equation*}
\vartheta : \ovr[\mathfrak{ug}] \; {\Join}  \;\ovl[\mathfrak{ug}]
\longrightarrow
\ovr[\mathfrak{ug}] \; {\Join}  \;\ovl[\mathfrak{ug}],
\end{equation*}
\noindent
defined by
\begin{equation*}
\vartheta (J + K) = J - K,\quad \forall J \in \mathfrak{j},
\quad \forall K \in \mathfrak{k}.
\end{equation*}
\noindent
$\mathfrak{j}$ is the set of fixed points of $\vartheta$. It is 
unique (\cite{He}, Chapter IV, \S 3, Proposition 3.5).\\
\indent
The pair $(\ovr[\mathfrak{ug}] \; {\Join}  \;\ovl[\mathfrak{ug}],
\; \vartheta)$ is an orthogonal symmetric Lie algebra
(\cite{He}, Chapter~IV, \S 3). There is a Riemannian symmetric
pair $(UG,\; SU_J(2))$ associated with $(\ovr[\mathfrak{ug}] \; 
{\Join}  \;\ovl[\mathfrak{ug}], \; \vartheta)$ so that the quotient
$UG/SU_J(2)$ is a complete locally symmetric Riemannian space.
Furthermore, its curvature corresponding to any $UG$-invariant
Riemannian structure is given by (\cite{He}, Chapter IV, \S 4, 
Theorem 4.2):
\begin{equation*}
R(K_{i_1}, K_{i_2})K_{i_3}= -[[K_{i_1}, K_{i_2}], K_{i_3}]
\quad \forall K_{i_1}, K_{i_2}, K_{i_3} \in \mathfrak{k}.
\end{equation*}
\noindent
Computing the sectional curvature we see that $R^{\text{sect}} 
\equiv 1$. Now a pedestrian version of the Sphere 
theorem~\cite{C-G} asseverates that a complete simply connected 
Riemannian manifold with $R^{\text{sect}} \equiv 1$ is isometric 
to a sphere of appropriate dimension. In our case the topological
condition is satisfied in view of Theorem~\ref{trivialhomotopy}.
\end{proof}

\section{Superspin structures}

\indent
The task ahead is clear: to find the relativistic transformation law 
for Dirac spinors. Any new group of symmetries (including the newly-minted 
$G$ of Section 2)  would still have to provide 
a bijective correspondence between two sets of solutions of the Dirac
equation - one being  the set of  original spinors, the other being the set
of transformed ones. At the same time this correspondence should not mess up
the spatial rotations of spinors. Last, but not least, the
resulting representation of $G$ has 
to be irreducible to ensure there is no mass splitting~\cite{O'R}.\\
\indent
To gain a better insight into the problem, prior to delving into the mire of
formulas, we discuss the concept of `free spin structure',
originally proposed by Plymen and Westbury~\cite{P-W}. This discussion 
might guide us towards a reasonable definition of the superspin structure. 
Thus let $M$ be a 4-dimensional smooth manifold with all the
obstructions to the existence of a Lorentzian metric vanishing 
(for instance, a parallelilazable $M$ would do). Let 
\begin{equation*}
\Lambda : \;\; \textnormal{Spin} (1, 3)^e \rightarrow SO(1, 3)^e
\end{equation*}
be the twofold covering epimorphism of Lie groups. A free spin structure
on $M$ consists of a principal bundle $\zeta : \Sigma \rightarrow M$ with
structure group $\textnormal{Spin} (1, 3)^e$ and a bundle map
$\widetilde{\Lambda}: \Sigma \rightarrow \mathcal{F} M$ into the bundle
of linear frames for $TM$, such that
\begin{equation*}
\widetilde{\Lambda} \circ {\widetilde{R}}_S = 
{\widetilde{R}}'_{\iota \circ \Lambda (S)} \circ \widetilde{\Lambda}
\;\;\; \forall S \in \textnormal{Spin} (1, 3)^e,
\end{equation*}
\begin{equation*}
{\zeta}' \circ \widetilde{\Lambda} = \zeta,
\end{equation*}
${\widetilde{R}}$ and ${\widetilde{R}}'$ being the canonical right actions
on $\Sigma$ and $\mathcal{F} M$ respectively, $\iota : SO(1,3)^e \rightarrow
GL(4, \mathbb{R})$ the natural inclusion of Lie groups, and ${\pi}' :
\mathcal{F} M \rightarrow M$ the canonical projection. The map 
$\widetilde{\Lambda}$ is called a spin-frame on $\textnormal{Spin}(1, 3)^e$.
This definition of a spin structure induces metrics on $\Sigma$. Indeed,
given a spin-frame $\widetilde{\Lambda}: \Sigma \rightarrow \mathcal{F} M$,
a dynamic metric $g_{\widetilde{\Lambda}}$ is defined to
be the metric that ensures orthonormality of all frames in 
$\widetilde{\Lambda}(\Sigma) \subset \mathcal{F} M$. It should be
emphasized that within the
Plymen and Westbury's formalism the metrics are built \textit{a posteriori},
after a spin-frame has been set by the field equations.\\
\indent
There is no way to extrapolate the above definition onto our framework
because our group in its present incarnation
does not act on any 4-dimensional manifold. However, 
their idea of carving a metric out of the spin structure permits a
not-so-literal generalization. A principal connection on the  $UG$
bundle over the physical space-time would quantify the amount by which a 
frame deviates from the standard Lorentz frame. Then there is a metric that
compensates for the deviation in such a manner as to appear to an observer
dwelling in that frame to be the standard Lorentz metric. To preserve the
commutation relations among the impulse operators we must insist on the
metric being flat. This, in turn, mandates the following extension of the
Einstein's Equivalence Principle: locally every noninertial frame is 
equivalent to a metric. There are familiar rotating or accelerating frames, 
entailing curved metrics. We postulate, that, in addition to those frames, 
some purely quantum noninertial frames are equivalent to flat but 
nonetheless nonstandard metrics. Unlike rotating and accelerating frames
however, the $G$-frames are globally 
equivalent to some nonstandard flat metrics. \\
\indent
We cannot eschew the representation of $G$ on Diff(${\mathbb{R}}^4)$. 
The presence of virtual frames effectively kills any chance of representing 
the group solely by inertial frames. \\
\indent
Now we set out to demonstrate that our program, spelled out above, is
viable. Consider the natural inclusions of Lie groups
\begin{equation}
\iota : UG \hookrightarrow GL(4, \mathbb{C}), \quad
\iota : \text{Spin}(1, 3)^e \hookrightarrow GL(4, \mathbb{C}).
\end{equation}
\noindent
Their images inside $GL(4, \mathbb{C})$ intersect:
\begin{equation}\label{properspin}
\iota (UG) \cap \iota (\text{Spin}(1, 3)^e)=SU_J(2).
\end{equation} 
\noindent
Because of~\eqref{properspin}, the set 
\begin{equation}
{\text{Ad}}_{\iota (UG)}(\iota (\text{Spin}(1, 3)^e))=
\coprod_{u \in UG} u\text{Spin}(1, 3)^e)u^H, 
\end{equation}
the disjoint union of conjugates of $\text{Spin}(1, 3)^e)$,
has the same cardinality as the set of 
all boosts in $UG$. Similarly, there is the natural inclusion
\begin{equation}
\iota :\quad SO(4) \hookrightarrow GL(4, \mathbb{R}).
\end{equation}
\noindent
The set ${\text{Ad}}_{\iota (SO(4))}(\iota (\text{SO}(1, 3)^e))$ is
homeomorphic to $SO(4)/SO(3) \cong {\mathbb{S}}^3$. Combining this with
Theorem~\ref{s3theorem} we arrive at two strings of relations running
parallel:
\begin{equation*}
\begin{CD}
\text{Ad}_{\iota (UG)}(\iota (\text{Spin}(1,3)^e)) @=
UG/SU_J (2) @>{\cong}>> {\mathbb{S}}^3\\ 
@.             @.                 @|\\
\text{Ad}_{\iota (SO(4))}(\iota (\text{SO}(1,3)^e))
@= SO(4)/SO(3) @>{\cong}>>  {\mathbb{S}}^3
\end{CD}
\end{equation*}
The double horizontal lines indicate set-theoretic bijective 
correspondences, the upper $\cong$ is an isometry, the lower one 
is a diffeomorphism. Furthermore, the diagram below commutes and
\textit{de facto} defines the superspin structure as conjugation of 
the free spin structure by the elements of $UG$.
\begin{equation*}
\begin{CD}
{\vec{\partial}}@>{UG-\text{connection}}>>{E\vec{\partial} + K}\\
@VVV                                                 @VVV\\
{{\text{Spin}(1, 3)}^e} @>{\text{Ad}}_{\iota (UG)}>>
{e^{iK}{\text{Spin}(1, 3)^e}}e^{-iK}\\
@VVV                                                 @VVV\\
{\widetilde{\Lambda}} @>{{\text{Ad}}_{\iota (SO(4))}}>>
{o\widetilde{\Lambda}o^T}\\
@VVV                                                 @VVV\\
g_{\widetilde{\Lambda}} @>>>  og_{\widetilde{\Lambda}}o^T 
\end{CD}
\end{equation*}
\noindent
Thus the superspin structure is a way to link groups $UG$-conjugate to 
$\text{Spin}(1, 3)^e$ inside $GL(4, \mathbb{C})$, with those 
$SO(4)$-conjugate to $SO(1,3)^e$ inside $GL(4, \mathbb{R})$. All 
conceivable superspin structures are parameterized by the elements of
$\text{Diff}({\mathbb{S}}^3)$. In particular, they can be bunched 
together into equivalence classes parameterized by 
${\pi}_3({\mathbb{S}}^3) = \mathbb{Z}$.\\
\indent
The reason our definition has some nontrivial content is, the group
$\text{Spin}(1, 3)^e$ features two inequivalent representations of
$SO(1,3)^e$ - (1/2, 0) and (0, 1/2). Had there been two equivalent
ones, the set ${\text{Ad}}_{\iota (UG)}(\iota (...))$ would have
consisted of only one element and the superspin structures would
have been reduced to the free spin structures.

\section{Relativistic coinvariance}
\indent
With the superspin structure in place we now nail down the particulars.
Instead of the standard quantum field theory substitution
\begin{equation}\label{standardQFT}
p_{\mu} \longrightarrow i{\partial}_{\mu},
\end{equation}
\noindent
we employ the rule
\begin{equation}\label{nabla}
p_{\mu} \longrightarrow i{\nabla}_{\mu}({\alpha}) 
\overset{\text{def}}{=} i({\varepsilon}^{\nu}_{\mu}(\alpha)
{\partial}_{\nu} + i{\kappa}_{\mu}^a (\alpha) K_a),
\end{equation}
\noindent
$K_a \in \mathfrak{k}$, ${\kappa}_{\mu}^a (\alpha)$ being a 
superspinor potential, chosen to make $i{\nabla}_{\mu}({\alpha})$
a purely imaginary operator. ${\nabla}_{\mu}({\alpha})$ qualifies 
as a $UG$-connection on the principal $UG$-bundle over the physical
space-time. Possibly, ${\kappa}_{\mu}^a (\alpha)$'s are functions
of the base space coordinates. The case of the flat space-time
can be elaborated at this point.
Assuming flatness, ${\kappa}_{{\mu}_l}^a 
(\alpha)$ may depend only on $x^0$ and $x^{{\mu}_l}$ to properly
convey the essence of the boost. Therefore, for a pure boost, only
two of four ${\kappa}_{\mu}^a (\alpha)$'s are nonzero for a fixed $a$;
of those, one is ${\kappa}_0^a (\alpha)$. An additional restriction
is entailed if we insist upon the Schr\H{o}dinger representation being
valid: $[p_{\mu}, p_{\nu}]=0$. To that end we need
\begin{equation}
({\varepsilon}^{0}_{0}(\alpha)
{\partial}_{0}{\kappa}_{{\mu}_l}^a (\alpha)+
{\varepsilon}^{{\mu}_l}_{0}(\alpha)
{\partial}_{{\mu}_l}{\kappa}_{{\mu}_l}^a (\alpha))-
({\varepsilon}^{0}_{{\mu}_l}(\alpha)
{\partial}_{0}{\kappa}_{0}^a (\alpha)+
{\varepsilon}^{{\mu}_l}_{{\mu}_l}(\alpha)
{\partial}_{{\mu}_l}{\kappa}_{0}^a (\alpha)) =0,
\end{equation}
\begin{equation}
[{\kappa}_{0}^1 (\alpha)K_1 + {\kappa}_{0}^2(\alpha)K_2 +
{\kappa}_{0}^3(\alpha)K_3, {\kappa}_{{\mu}_l}^1 
(\alpha)K_1 + {\kappa}_{{\mu}_l}^2(\alpha)K_2 +
{\kappa}_{{\mu}_l}^3(\alpha)K_3] = 0.
\end{equation}
\indent
Now by virtue of $[\mathfrak{j}, \mathfrak{k}] = \mathfrak{k}$,
for every space direction the corresponding boost must be obtainable
via some $SU_J(2)$ action on~\eqref{nabla}. That action ought to be 
linear to be truly spinorial:
\begin{alignat}{2}\label{su2action}
U{\gamma}^{\mu}{\nabla}_{\mu}U^H &=U{\gamma}^{\mu}U^H
{\varepsilon}^{\nu}_{\mu}{\partial}_{\nu} + i{\kappa}_{{\mu}}^a
U{\gamma}^{\mu}U^HUK_aU^H& &\quad\\
&=M^{\mu}_{\eta}{\gamma}^{\eta}{\varepsilon}^{\nu}_{\mu}{\partial}_{\nu}
+M^{\mu}_{\eta}{\gamma}^{\eta}i{\kappa}_{{\mu}}^a r^n_a K_n & &
\quad \text{by}\;[\mathfrak{j}, \mathfrak{k}] = \mathfrak{k}.\notag
\end{alignat}
\noindent
Here $M^{\mu}_{\eta}$'s realize an $SO(3)$ transformation 
$(U \in SU_J(2))$, which is
at its most transparent if ${\gamma}^{0}$ is diagonal. 
As for $r_a^n$'s, they determine how the potentials behave:
\begin{equation}
{\Tilde{\kappa}}^a_{\mu}={\kappa}_{{\mu}}^1 r^a_1 +
{\kappa}_{{\mu}}^2 r^a_2 + {\kappa}_{{\mu}}^3 r^a_3, \quad {\text{and}}
\end{equation}
\begin{equation}
{|r^a_1|}^2 +{|r^a_2|}^2 +{|r^a_3|}^2=1,\quad a=\{1,2,3\}.
\end{equation} 
\indent
In order for us to express ${\kappa}_{\mu}^a (\alpha)$ explicitly as 
functions of $\alpha$, we have to introduce the concept of relativistic
coinvariance.
We define the relativistic coinvariance to be a twofold property of our
mathematical formalism; that the impulse operators
transform via a  principal $UG$ connection,
and, at the same time, this connection complies with the relativistic 
invariance law
\begin{equation}\label{coinvariance}
\boxed{{\Tilde{p}}^{\mu}{\Tilde{p}}_{\mu}= g^{\nu \lambda}({\alpha})
{\nabla}_{\nu}({\alpha}){\nabla}_{\lambda}({\alpha})
\overset{\text{def}}{=} g^{\nu \lambda}(0)
{\partial}_{\nu}{\partial}_{\lambda}= p^{\mu}p_{\mu}}
\end{equation}
\noindent 
translating to some algebraic relations between ${\kappa}_{\mu}$'s.
A pure boost is best exemplified by the boost in the $x^{{\mu}_l}$-direction.
The metric deforms as follows:
\begin{equation}
g^{00}=\cos \alpha,\quad g^{{\mu}_l{\mu}_l}=-\cos \alpha,
\quad g^{0{\mu}_l}=g^{{\mu}_l0}= \sin \alpha.
\end{equation} 
For that particular transform we have
\begin{align}
({{\kappa}_{0}}^2 - {{\kappa}_{{\mu}_l}}^2)\cos \alpha +
2{\kappa}_{0}{\kappa}_{{\mu}_l} \sin \alpha& =\phantom{-} 0,\\
({{\varepsilon}^{0}_{0}}^2 -{{\varepsilon}^{0}_{{\mu}_l}}^2)
\cos \alpha + 2{\varepsilon}^{0}_{0}{\varepsilon}^{0}_{{\mu}_l}
\sin \alpha& =\phantom{-}1,\\
({{\varepsilon}^{{\mu}_l}_{0}}^2 -
{{\varepsilon}_{{\mu}_l}^{{\mu}_l}}^2)\cos \alpha + 
2{\varepsilon}^{{\mu}_l}_{0}{\varepsilon}_{{\mu}_l}^{{\mu}_l}\sin \alpha&=-1,\\
({\varepsilon}^{0}_{0}{\varepsilon}^{{\mu}_l}_{0} - {\varepsilon}^{0}_{{\mu}_l}
{\varepsilon}_{{\mu}_l}^{{\mu}_l})\cos \alpha +
({\varepsilon}^{0}_{0}{\varepsilon}_{{\mu}_l}^{{\mu}_l} +
{\varepsilon}^{{\mu}_l}_{0}{\varepsilon}^{0}_{{\mu}_l})\sin \alpha&=\phantom{-}0.
\end{align}
\noindent
Without the no torsion assumption (which may be extraneous in the curved
space-time), the last equation splits into
\begin{equation}
\begin{cases}
({\varepsilon}^{0}_{0}{\varepsilon}^{{\mu}_l}_{0}-{\varepsilon}^{0}_{{\mu}_l}
{\varepsilon}_{{\mu}_l}^{{\mu}_l})\cos \alpha =0,\\
({\varepsilon}^{0}_{0}{\varepsilon}_{{\mu}_l}^{{\mu}_l} +
{\varepsilon}^{{\mu}_l}_{0}{\varepsilon}^{0}_{{\mu}_l})\sin \alpha\, =0.
\end{cases}
\end{equation}
\noindent
These boosts are not linear, generally speaking, yet with all the 
above-listed constraints the remaining arbitrariness is considerably 
less than the arbitrariness of internal symmetries and gauge
transformations. It is reflected in the Lagrangian being 
given by a familiar expression~\cite{L}:
\begin{equation}
{\mathcal{L}}_D = \frac{i}{2}({\Psi}^{\dag}{\gamma}^{\mu}
{\nabla}_{\mu}{\Psi}- {\nabla}_{\mu}{\Psi}^{\dag}{\gamma}^{\mu}
{\Psi}- m{\Psi}^{\dag}{\Psi}).
\end{equation}
\noindent
The one crucial distinction we want to make is that in the present
context, ${\nabla}_{\mu}$'s  stand for components of a principal
connection, rather than the metric connection.\\
\indent
If instead of~\eqref{standardQFT} the minimal substitution
\begin{equation}
p_{\mu} - e{\mathbb{A}}_{\mu}
\longrightarrow i{\partial}_{\mu} - e{\mathbb{A}}_{\mu}
\end{equation}
\noindent
is used, we set
\begin{equation}
p_{\mu} - e{\mathbb{A}}_{\mu}\longrightarrow 
i{\nabla}_{\mu}({\alpha}) - e{\mathbb{A}}_{\mu}(\alpha)
\overset{\text{def}}{=}i( {\varepsilon}^{\nu}_{\mu}(\alpha)
{\partial}_{\nu} + {\kappa}_{\mu}^a (\alpha) K_a)
- e{\mathbb{A}}_{\mu}(\alpha).
\end{equation}
\noindent
The coinvariance condition~\eqref{coinvariance} then becomes
\begin{equation}
\begin{split}
({\Tilde{p}}^{\mu}-e{\Tilde{{\mathbb{A}}}}^{\mu})
({\Tilde{p}}_{\mu}-e{\Tilde{{\mathbb{A}}}}_{\mu})
&=g^{\nu \lambda}({\alpha})
({\nabla}_{\nu}({\alpha})- e{\Tilde{{\mathbb{A}}}}_{\nu}(\alpha))
({\nabla}_{\lambda}({\alpha})-
e{\Tilde{{\mathbb{A}}}}_{\lambda}(\alpha))\\
&= g^{\nu \lambda}(0)({\partial}_{\nu}- e{\mathbb{A}}_{\nu}(0))
({\partial}_{\lambda}- e{\mathbb{A}}_{\lambda}(0))\\
&=(p^{\mu}-e{\mathbb{A}}^{\mu})(p_{\mu}-e{\mathbb{A}}_{\mu}).
\end{split}
\end{equation}
\noindent 
Superspinors are invariant with respect to the gauge 
transformations:
\begin{equation}
\Tilde{\Psi}(x, \alpha) = e^{if(x)}{\Psi}(x, \alpha),
\end{equation}
\begin{equation}
{\Tilde{{\mathbb{A}}}}_{\nu}(x,\alpha) =
{\mathbb{A}}_{\nu}(x,\alpha)- e^{-1}{\varepsilon}^
{\mu}_{\nu}(\alpha){\partial}_{\mu} f(x),
\end{equation}
\noindent
where $f(x)$ is an arbitrary real function of space-time coordinates.

\section{Solutions}

\indent
The modification of the Dirac equation effected by our prescription
${\partial}_{\mu} \Longrightarrow {\nabla}_{\mu}$ leads to
\begin{equation}\label{Dirac1}
(i {\gamma}^{\mu}{\nabla}_{\mu} - m)\Psi = 0,
\end{equation}
\begin{equation}\label{Dirac2}
(i {\gamma}^{\mu}{\nabla}_{\mu} - m){\Phi} = 0,
\end{equation}
\noindent
corresponding to the ordinary positive and negative energy spinors: 
\begin{align}
({\gamma}^{\mu}p_{\mu} - m)&w = 0,\\
({\gamma}^{\mu}p_{\mu} + m)&u\, = 0.
\end{align}
\indent
We confine ourselves to a prototypical case - that of a boost in the
$x^3$ direction. Specifically,
\begin{align}
{\nabla}_{0}&={\varepsilon}^{0}_{0}(\alpha){\partial}_0 + 
{\varepsilon}^{3}_{0}(\alpha){\partial}_3 + i{\kappa}_0(\alpha)K_3,\\
{\nabla}_{3}&={\varepsilon}^{0}_{3}(\alpha){\partial}_0 + 
{\varepsilon}^{3}_{3}(\alpha){\partial}_3 + i{\kappa}_3(\alpha)K_3,\\
{\nabla}_{1} &= {\partial}_1,\\
{\nabla}_{2} &= {\partial}_2.
\end{align}
\noindent
All other free superspinors can be obtained from 
these ones via the linear $SU_J(2)$ transformations~\eqref{su2action}. 
We look for plane-wave particle and antiparticle spinors (\cite{D}, 
Chapter XI, \S\S 70-73) of the form 
\begin{align}
\Psi(\alpha)&= w(\alpha)e^{-i(s_0(\alpha)x^0 +s_3(\alpha)x^3)},\\
\bar{\Phi}(\alpha)&=\bar{u}(\alpha)e^{-i(s_0(\alpha)x^0 +s_3(\alpha)x^3)},
\end{align}
subject to the relativistic impulse condition 
${s_0}^2(\alpha)-{s_3}^2(\alpha)=m^2$. This is a \textit{conditio sine
qua non} because every component ${\Psi}^l(\alpha)$ of
$\Psi(\alpha)$ and ${\bar{\Phi}}^l(\alpha)$ must satisfy the 
Klein-Gordon equation 
\begin{align}
(\square + m^2){\Psi}^l(\alpha)&=0,\\
(\square + m^2){\bar{\Phi}}^l(\alpha)&=0.
\end{align}
\noindent
In the standard representation
\begin{equation}
{\gamma}^0 =
\begin{bmatrix}
1&\phantom{-}0\\
0&-1
\end{bmatrix},\quad {\gamma}^{i} =
\begin{bmatrix}
0&-{\sigma}_i\\
{\sigma}_i&\phantom{-}0
\end{bmatrix},
\end{equation}
\noindent
the equations \eqref{Dirac1} and \eqref{Dirac2} yield 
the following matrix:
\begin{equation}
\begin{bmatrix}
{\varepsilon}_{0}(\alpha)-m(\alpha)
&0&-{\varepsilon}_{3}(\alpha)-{\kappa}_{0}(\alpha)&0\\
0&{\varepsilon}_{0}(\alpha)-m(\alpha)
&0&{\varepsilon}_{3}(\alpha)+{\kappa}_{0}(\alpha)\\
{\varepsilon}_{3}(\alpha)-{\kappa}_{0}(\alpha)
&0&-{\varepsilon}_{0}(\alpha)-m(\alpha)&0\\
0&-{\varepsilon}_{3}(\alpha)+{\kappa}_{0}(\alpha)
& 0&-{\varepsilon}_{0}(\alpha)-m (\alpha) 
\end{bmatrix}, 
\end{equation}
\noindent
where the entries are
\begin{align}
{\varepsilon}_{0}(\alpha)&={\varepsilon}^{0}_{0}(\alpha)s_0 + 
{\varepsilon}^{3}_{0}(\alpha)s_3,\\
{\varepsilon}_{3}(\alpha)&={\varepsilon}^{0}_{3}(\alpha)s_0 + 
{\varepsilon}^{3}_{3}(\alpha)s_3,\\
m(\alpha)&= m +{\kappa}_{3} (\alpha).  
\end{align}
\indent
Its rank has to be 2 for all values of $\alpha$, thus 
constraining ${\kappa}_0(\alpha)$ and ${\kappa}_3(\alpha)$:
\begin{equation}
{{\varepsilon}_{0}}^2(\alpha) -  {{\varepsilon}_{3}}^2(\alpha)=
(m +{\kappa}_{3}(\alpha))^2 -{{\kappa}_0}^2(\alpha).
\end{equation}
\noindent
At last, the proper role of ${\kappa}_{\mu}(\alpha)$'s is
revealed: they make mass into a quantity that serves as
such in noninertial frames. To keep this mass term position-invariant 
we must ensure that ${\partial}_{\nu} {\kappa}_{\mu}(\alpha) =0$. 
The solutions are
\begin{equation}
w^{(1)}(\alpha)=
\begin{bmatrix}
{\varepsilon}_{0}(\alpha)+m +{\kappa}_{3} (\alpha)\\
0\\
{\varepsilon}_{3}(\alpha)-{\kappa}_0(\alpha)\\
0
\end{bmatrix},\quad
w^{(2)}(\alpha)=
\begin{bmatrix}
0\\
{\varepsilon}_{0}(\alpha)+m +{\kappa}_{3} (\alpha)\\
0\\
-{\varepsilon}_{3}(\alpha)+{\kappa}_0(\alpha)
\end{bmatrix},
\end{equation}
\begin{equation}
u^{(1)}(\alpha)=
\begin{bmatrix}
{\varepsilon}_{3}(\alpha) +{\kappa}_{0} (\alpha)\\
0\\
{\varepsilon}_{0}(\alpha) -m -{\kappa}_3(\alpha)\\
0
\end{bmatrix},\quad
u^{(2)}(\alpha)=
\begin{bmatrix}
0\\
-{\varepsilon}_{3}(\alpha)-{\kappa}_{0} (\alpha)\\
0\\
{\varepsilon}_{0}(\alpha)-m -{\kappa}_3(\alpha)
\end{bmatrix}.
\end{equation}
\noindent
The crucial values are $\alpha = \{0, {\pi}/{2}, \pi, 
{3\pi}/{2} \}$. The first two: 
\begin{equation*}
\begin{aligned}
&{\varepsilon}^{0}_{0}(0)={\varepsilon}^{3}_{3}(0)=1\\ 
&{\varepsilon}^{3}_{0}(0)={\varepsilon}^{0}_{3}(0)=0\\ 
&{\kappa}_{3}(0)={\kappa}_0(0)=0  
\end{aligned}\qquad \;\;\;\;
\begin{aligned}  
&{\varepsilon}^{0}_{0}({\pi}/{2})=
{\varepsilon}^{3}_{3}({\pi}/{2})=\sqrt{2}/{2}\\
&{\varepsilon}^{3}_{0}({\pi}/{2})=
-{\varepsilon}^{0}_{3}({\pi}/{2})=-\sqrt{2}/{2}\\        
&\lim_{{\alpha} \to \frac{\pi}{2}}
((m +{\kappa}_{3}(\alpha))^2 -{{\kappa}_0}^2(\alpha))= \infty
\end{aligned}
\end{equation*}
\noindent
mirror the second two:
\begin{align*}
&{\varepsilon}^{0}_{0}(\pi)={\varepsilon}^{3}_{3}(\pi)=0\\ 
&{\varepsilon}^{3}_{0}(\pi)={\varepsilon}^{0}_{3}(\pi)=-1\\ 
&(m -{\kappa}_{3}(\pi))^2 -{{\kappa}_0}^2(\pi)=-m^2  
\end{align*}
\begin{align*}  
&{\varepsilon}^{0}_{0}({3\pi}/{2})=
{\varepsilon}^{3}_{3}({\pi}/{2})=\sqrt{2}/{2}\\
&{\varepsilon}^{3}_{0}({3\pi}/{2})=-
{\varepsilon}^{0}_{3}({3\pi}/{2})=\sqrt{2}/{2}\\        
&\lim_{{\alpha} \to \frac{3\pi}{2}}
((m +{\kappa}_{3}(\alpha))^2 -{{\kappa}_0}^2(\alpha))= -\infty.
\end{align*}
\indent
An unexpected relation between particles and antiparticles emerges:
\begin{equation}\label{super1}
{\Psi}^{(1)}(\alpha)=-{\bar{\Phi}}^{(1)}(\alpha + \pi),
\end{equation}
\begin{equation}\label{super2}
{\Psi}^{(2)}(\alpha)=\;\;\,{\bar{\Phi}}^{(2)}(\alpha + \pi).
\end{equation}
\noindent
Verbally, virtualization and actualization in the context of 
space-time superspinor transformations occur only in conjunction
with the charge conjugation. According to~\eqref{super1},
~\eqref{super2}, electrons are virtual positrons and
\textit{vice versa}. The superspinor formalism forestalls their
leaving the mass surface, yet recognizes the difference between
actual and virtual charged spin 1/2 particles associated with a
frame. There is no question of moving with a superlight speed, 
for no particle would retain its original identity.
In this brave new world electrons and positrons are just particular
values of the superspinor wave function. Also, the particle-antiparticle
symmetry hypothesis (\cite{D}, Chapter XI, \S 73) is ultimately
vindicated, since the vacuum is filled with all kinds of negative 
energy superspinors, and must be electrically indefinite. Needless 
to say, the energy-impulse is commensurate with the frame rapidity, but 
the act of virtualization $({\Psi}(\alpha)\mapsto 
{\Psi}(\alpha + \pi))$ must preserve it:
\begin{equation}\label{superenergy}
{s}_{\mu}(\alpha) ={s}_{\mu}(\alpha + \pi).
\end{equation}

\section{Superspinor statistics}
\indent
Even though the results of the previous section were obtained for a
specialized $UG$-transformation, they obviously remain true for all
superspinors. Thus $\alpha$ serves as a universal boost parameter.\\
\indent
An arbitrary solution allows plane-wave  
decompositions (\cite{R}, Chapter 4, \S 4.3):
\begin{align}\label{Fourier}
\Psi (x)& = \int \frac{d^3 s}{(2\pi)^3}\frac{m}{s_0}\sum_{l=1,2}
[{\mathcal{B}}_l(s)w^l(s)e^{-isx} +{\mathcal{D}}_l^H(s)u^l(s)e^{isx}],\\
\bar{\Psi}(x)& = \int \frac{d^3 s}{(2\pi)^3}\frac{m}{s_0}\sum_{l=1,2}
[{\mathcal{B}}_l^H(s)\bar{w}^l(s)e^{isx} 
+{\mathcal{D}}_l(s)\bar{u}^l(s)e^{-isx}].
\end{align}
\noindent
In these formulas ${\mathcal{B}}_l(s)$ and ${\mathcal{D}}_l(s)$ are 
viewed as linear operators, not just coefficients, 
and would have to be interpreted as such.
Flipping \eqref{Fourier} and using \eqref{superenergy}, we get
\begin{equation}\label{superoperators}
{\mathcal{B}}_l(s(\alpha))=(-1)^{l}{\mathcal{D}}_l^H(s(\alpha)).
\end{equation}
\noindent
It stands to reason, that, essentially, creating a
particle is equivalent to annihilating an antiparticle.\\
\indent
By virtue of~\eqref{superoperators}, superspinors 
entail the following anticommutators:
\begin{align}
\{ {\mathcal{D}}_l^H(s(\alpha)),{\mathcal{B}}_n(s'(\alpha))\}&=
\pm \{ {\mathcal{B}}_l(s(\alpha)), {\mathcal{B}}_n(s'(\alpha))\}=
\{ {\mathcal{B}}_l(s(\alpha)), {\mathcal{D}}_n^H(s'(\alpha))\},\\
\{ {\mathcal{B}}_l^H(s(\alpha)), {\mathcal{D}}_n(s'(\alpha))\}&=
\pm \{ {\mathcal{D}}_l(s(\alpha)), {\mathcal{D}}_n(s'(\alpha))\}=
\{ {\mathcal{D}}_l(s(\alpha)), {\mathcal{B}}_n^H(s'(\alpha))\}.
\end{align}
\noindent
Whenever ${s}(\alpha)\ne {s'}(\alpha)$, the only way for the
left- and right-hand side anticommutators to be equal is to vanish,
because creating a particle with impulse $s$ combined with
annihilating an antiparticle with impulse $s'$ is fundamentally
different from creating a particle with impulse $s'$ combined with
annihilating an antiparticle with impulse $s$. Now the above
anticommutators must continuously depend on the impulse. Therefore
\begin{equation}
\{ {\mathcal{B}}_l(s(\alpha)), {\mathcal{B}}_n(s'(\alpha))\}=
\{ {\mathcal{D}}_l(s(\alpha)), {\mathcal{D}}_n(s'(\alpha))\}=0, 
\quad \forall s, s'.
\end{equation}
\noindent
Hence ${\mathcal{B}}_l(s(\alpha)) {\mathcal{B}}_l(s(\alpha))=0$, 
and furthermore ${\mathcal{B}}_l(s(\alpha)) 
{\mathcal{B}}_l(s(\alpha))|\quad \rangle =0$. This in
fact says that two superspinors with the definite impulse $s$,  
an identical spin, and an identical charge cannot be in the same
state. We conclude that for superpinors, the Fermi-Dirac
statistics comes about as a direct consequence of the 
relativistic coinvariance, whereas the conventional Dirac spinors 
need additional anticommutator relations - the Jordan-Wigner 
postulates (\cite{R}, Chapter 4, \S 4.3). 

\section{Twin paradox for superspinors}

\indent
A simple way to determine whether the superspinor model has any 
semblance to the real world is to conduct an experiment in the setting 
similar to that of the twin paradox experiment.
Let us let one local frame move, while the other be still. Let there be
an electromagnetic field expressible in the moving frame as
$-e {\mathbb{A}}_{\mu}$. At the exact moment these two frames coincide 
in space, let that exact location be bombarded  with
a gravitational wave decomposable into two pieces: $\Gamma = 
({\kappa}^a_{\mu}K_a + \Re (e {\mathbb{A}}_{\mu}))$. Then 
$\pm {\kappa}^a_{\mu}K_a$ and $\pm \Re(e {\mathbb{A}}_{\mu})$
cancel, and electrons (in fact, any massive spin 1/2 particles participating 
in the electromagnetic interactions) would behave differently in these two 
frames. Now apply a uniform gravitational wave over a region in space. Let 
the moving frame be associated with a spacecraft. When it finally returns to 
the location of the resting local frame, the differences in the electron
superspinors congeal and become absolute. More specifically, the electrons
on the spaceship would be impervious to the action of the uniform
gravitational wave - a levitation of sorts.\\
\indent
This can be seen as a mirror image of the Aharonov-Bohm~\cite{A-B}
phenomenon. Indeed, a change in the fermion field triggered on the moving 
spaceship by the uniform gravitational wave (which essentially is a
space-time deformation, albeit not necessarily topologically 
nontrivial one) causes changes in the 
electromagnetic field. Globally, there is an interdependency between 
massive spin 1/2 particles and electromagnetic fields. Direct interaction 
cannot account for all of that interdependency. We would like to call it 
the Aharonov-Bohm symmetry. Its secret is hidden deep in the topology and 
small-scale structure of the space-time. We can only speculate that such 
conundrums as the self-action of the electric field of an electron, or the 
electromagnetic mass will find some measure of elucidation within a framework 
encompassing the Aharonov-Bohm symmetry.\\

\pagebreak

\end{document}